\documentclass{article}
\usepackage{spconf,amsmath,graphicx}
\usepackage{multirow}
\usepackage{url}

\title{Towards Developing State-of-the-Art TTS Synthesisers for 13 Indian Languages with Signal Processing aided Alignments}
\name{Anusha Prakash, S Umesh, Hema A Murthy}
\address{Indian Institute of Technology Madras, India}

\copyrightnotice{979-8-3503-0689-7/23/\$31.00~\copyright2023 IEEE}

\begin{document}
%

\maketitle
\begin{abstract}
End-to-end (E2E) systems synthesise high-quality speech, but this typically requires a large amount of data. As E2E synthesis progressed from Tacotron to FastSpeech2, it became evident that features representing prosody, particularly sub-word durations, are important for error-free synthesis. Variants of FastSpeech use a teacher model or forced alignments for training. This paper uses signal processing cues in tandem with forced alignment to produce accurate phone boundaries for the training data. As a result of better duration modelling, good-quality synthesisers are developed. Evaluations indicate that systems developed using the proposed signal processing-aided approach are better than systems developed using other alignment approaches, especially in low-resource scenarios. Our systems also outperform the existing best TTS systems available for 13 Indian languages.
\end{abstract}
\section{Introduction}
\label{sec:intro}

India has a wide linguistic diversity with 1369 languages, including 23 official languages \cite{censusIndia2011}. Of its one billion+ population, only $74.04\%$ is literate. Poor literacy levels and limited or no proficiency in English underscore the development of good-quality Indic speech synthesis systems to better engage the general public. This task is challenging given that most Indian languages have limited or no resource availability. In this work, we develop good quality end-to-end (E2E) text-to-speech (TTS) systems for Indian languages by seamlessly integrating signal processing cues in deep learning techniques. Specifically, the focus is on improving the duration prediction (and thereby the synthesis quality) of the E2E TTS systems by correcting the phone alignments of the training data.




The E2E approach is the popular state-of-the-art speech synthesis paradigm due to its ease in training systems to obtain high-quality speech. The initial E2E systems were primarily attention-based, such as Tacotron \cite{tacotron}, Tacotron2 \cite{tacotron2}. The main goal of the attention module in TTS tasks is to learn the alignments between characters/phones and mel-spectrogram frames. The attention module learns soft alignments, in comparison to hard alignments used in traditional TTS approaches such as unit selection synthesis (USS) \cite{Hunt96} and hidden Markov model (HMM) based speech synthesis systems (HTS) \cite{tokuda}. The attention network is trained to enable duration prediction during synthesis.


One of the main drawbacks of attention-based networks is that the alignments may not be learnt correctly. Coupled with an auto-regressive decoder, the synthesised output is prone to errors, such as the insertion or deletion of phones. Hence, the focus of the E2E paradigm has shifted to improving duration prediction during synthesis. The duration information corresponding to the training data can be learnt in different ways. FastSpeech \cite{Fastspeech} uses alignments predicted by a teacher model. Some architectures, such as FastSpeech2 \cite{fastspeech2} and DurIAN \cite{durian_IS2020}, employ external aligners for this purpose. Other recent works, such as \cite{shih2021radtts, YasudaWY21, OneTTSAlignment_2022, lim22_interspeech, GlowTTS, VITS_2021} learn the alignments internally.


In the context of HMM-based systems, \cite{aswin14, BABY202010} have studied the effect of accurate alignments on synthesis quality and intelligibility, highlighting the importance of accurate boundaries for training. Current E2E TTS architectures employ machine learning based alignments for system building. Signal processing primarily depends on the acoustic characteristics of the speech signal and is agnostic to the transcriptions. Does combining their complementary features also help in E2E training, as already evidenced in the HTS and conventional neural network-based frameworks? Such a study is very important to produce good quality speech as duration is a vital prosody marker. We employ an external aligner, the hybrid segmentation (HS) algorithm, which combines signal processing cues in tandem with deep learning techniques \cite{BABY202010}, to obtain accurate alignments for the training data. We use the FastSpeech2 architecture \cite{fastspeech2}, and the HiFi-GAN v1 vocoder \cite{hifigan_NEURIPS2020} for E2E training\footnote{In this paper, the two-stage pipeline of generating mel-spectrograms and then reconstructing waveforms is also considered as E2E training.}.





The E2E system trained using the signal processing aided hybrid segmentation approach is referred to as the proposed system. The performance of the proposed system is compared with systems trained with different alignment techniques-- using a teacher model and Montreal forced aligner (MFA) \cite{MFA_interspeech17}. We also conduct experiments in a low-resource scenario, including a comparison with a direct text-to-wave VITS model \cite{VITS_2021}. Formal evaluations and qualitative observations indicate that the signal-processing aided system is comparable to or better than systems trained with purely machine learning-based alignments.



We also investigate how these trained systems compare with state-of-the-art TTS systems available for 13 Indian languages \cite{ai4bharat_TTS_2023} and evaluate system performance using subjective measures. This in itself is quite a challenge, as getting native listeners for each language is difficult. Most studies focus only on a few major languages and have many evaluators. In this work, we have tried our best to get as many evaluators as possible in each of the 13 languages. Subjective evaluations indicate an average preference of $62.63\%$ for the proposed systems over the systems of \cite{ai4bharat_TTS_2023}.

The rest of the paper is organised as follows. Section \ref{sec:related} reviews the related work. The baseline and proposed systems are presented in Section \ref{sec:baseline_proposed}. Associated experiments are described in Section \ref{sec:experiments}. The work is concluded in Section \ref{sec:conclusion}.
\section{Related work}
\label{sec:related}


This section presents recent literature focusing on duration modelling in E2E training. In FastSpeech \cite{Fastspeech}, duration information is obtained from a Transformer TTS \cite{transformer_TTS}, which is considered a teacher model. External aligners are used in a few papers-- MFA \cite{fastspeech2, MFA_interspeech17}, HMM-based \cite{parallelTac}, and connectionist temporal classification (CTC) based \cite{talknet_interspeech}. TTS systems trained in \cite{shih2021radtts, OneTTSAlignment_2022, lim22_interspeech, neuralHMMS_2022, VAE_2022} learn duration information internally using HMM-based approaches. Soft and hard alignments are learnt with monotonicity constraint in \cite{shih2021radtts, OneTTSAlignment_2022, lim22_interspeech}. Glow-TTS \cite{GlowTTS} uses normalizing flows and dynamic programming to determine the most probable monotonic alignments between text and the latent audio representation. Variational autoencoder with adversarial learning text-to-speech system (VITS) \cite{VITS_2021} also uses the monotonic alignment search (MAS) proposed in \cite{GlowTTS}. In \cite{PortaSpeech}, word-level hard alignments are obtained from an external aligner, and soft phone alignments are learnt using a word-to-phone attention network. A recently developed network called SoftSpeech \cite{softspeech_2022} proposes a soft length regulator for unsupervised duration modelling within the FastSpeech2 network. Among the presented literature, \cite{VAE_2022, softspeech_2022} demonstrate the capability of their TTS systems in low resource scenarios. 

 HMM-based alignments and MAS are both statistical-based approaches. In \cite{aswinThesis}, it is seen that forced alignment using HMMs does not always provide accurate alignments, especially for fricatives, affricates and nasals. The boundaries of these classes of sounds are refined using signal processing cues. Hence, in the current work, we use the hybrid segmentation algorithm \cite{BABY202010} to obtain accurate phone boundaries for the training data. Hybrid segmentation is an external aligner that combines the complementary features of signal processing and neural network-based techniques.

  A recently published paper \cite{ai4bharat_TTS_2023} has built good quality TTS systems for 13 Indian languages by exploring different state-of-the-art models employing different alignment techniques. This includes MAS in Glow-TTS \cite{GlowTTS} and VITS \cite{VITS_2021}, and the alignment learning framework in \cite{shih2021radtts, OneTTSAlignment_2022}. Evaluations show that the 2-stage pipeline (FastPitch \cite{fastpitch} + HiFiGAN \cite{hifigan_NEURIPS2020}) performs better than the direct end-to-end VITS model in terms of intelligibility. In the current work, we choose the FastSpeech2 network as the mel-spectrogram generation model as it provides more variance information with pitch and energy. On average, our proposed systems perform better than the best systems of \cite{ai4bharat_TTS_2023}.

    A recent work \cite{Alignment_ICASSP23} shows that small alignment errors (less than 75 msec) do not impact synthesis quality. But our experience with alignments and studies in \cite{fastspeech2} show that better alignments lead to better synthesis output.  
    
  

 
\section{Baseline and Proposed Systems}
\label{sec:baseline_proposed}
We present the baseline systems used in this work and describe in detail the hybrid HMM-GD-DNN segmentation (HS) approach, the alignment technique which we propose to be used for FastSpeech2. Then we briefly describe the E2E pipeline used.


\subsection{Baseline systems}

We consider the following baseline systems employing different alignment techniques:

\begin{enumerate}
    \item FastSpeech2 with teacher-student approach (TS): In the teacher-student approach, phone durations from an auto-regressive Tacotron2 teacher model (or Transformer network) are fed to FastSpeech2 model training. From a trained teacher model, encoder-decoder attention alignments are extracted for every $<$text, audio$>$ pair as described in \cite{Fastspeech}. 
    
    \item FastSpeech2 with Montreal forced aligner (MFA) \cite{MFA_interspeech17}: MFA is an open-source speech-text aligner that provides phone and word level boundaries. MFA performs triphone modelling and performs speaker adaptation to model inter-speaker differences. Models are trained in MFA using the Kaldi speech recognition toolkit \cite{Povey_ASRU2011_2011}. 
    
    \item VITS with monotonic alignment search (MAS): VITS is a direct E2E architecture \cite{VITS_2021} that uses a variational autoencoder to generate speech from text. VITS learns the phone alignments internally from the data using the monotonic alignment search (MAS) of Glow TTS \cite{GlowTTS}. MAS is a dynamic programming based approach to finding the optimal alignment between a speech waveform and its corresponding transcriptions. The alignments are restricted to non-skipping and monotonic. Training a VITS model is computationally intensive and requires a longer training time. Hence, a VITS model has been trained only in the low-resource scenario for performance comparison.
\end{enumerate}





\subsubsection{Hybrid HMM-GD-DNN segmentation (HS)}
Hybrid segmentation is an alignment technique that combines the complementary features of machine learning and signal processing-based approaches to generate accurate phone boundaries \cite{aswin14, BABY202010}. HMM-based forced alignment does not accurately model the location of phone boundaries. Hence, in \cite{aswin14}, these boundaries are corrected using signal processing-based cues. Specifically, a group delay (GD) based algorithm is used to obtain accurate syllable boundaries. However, the drawback of the GD-based technique is that it doesn't capture the correct number of syllable boundaries as it is agnostic to the text. Hence, spurious GD boundaries are estimated, and the GD boundary closest to an HMM boundary is considered the correct syllable boundary \cite{aswin14}. Then the phone boundaries are re-estimated within these syllable boundaries instead of re-estimating across the entire utterance.

Additionally, sub-band spectral flux (SBSF) is used as a cue for correcting boundaries of fricatives, affricates and nasals \cite{aswinThesis}. The boundaries of these sounds are characterised by significant spectral changes. Affricates and sibilant fricatives have high energy content in the higher frequency bands, while the energy content of nasals is more prominent in the lower frequency bands.

In \cite{BABY202010}, the accuracy of phone boundaries and the synthesis quality is compared across TTS systems trained with only deep neural network (DNN) alignments and with DNN alignments employing boundary correction. In the latter, the alignments obtained by the hybrid HMM-GD technique are considered initial alignments for DNN segmentation. Experiments show that the synthesis quality with boundary correction is better than with only DNN alignments. Motivated by this, we use the hybrid HMM-GD-DNN alignments for FastSpeech2 training and compare systems trained with the other machine learning based alignments discussed previously.  

\begin{figure}[!h]
 \centering
 \includegraphics[width = \linewidth, height = 3.5cm]{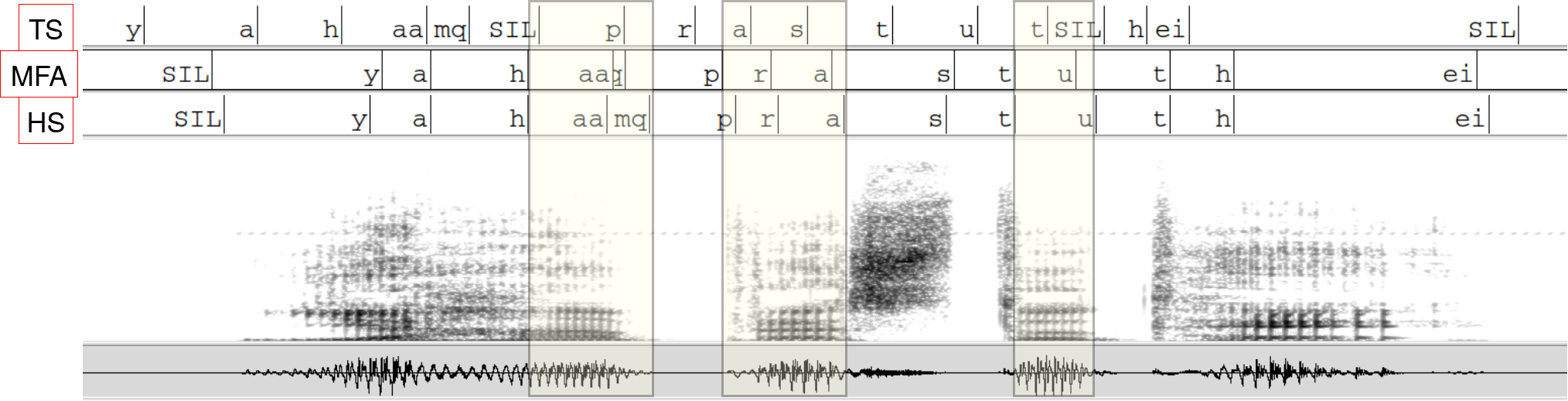}
   \caption{An example of a Hindi waveform (bottom panel), its spectrogram (fourth panel), and phone-level alignments obtained from different techniques (top 3 panels). TS: teacher-student approach, MFA: Montreal forced aligner, HS: hybrid segmentation. The highlighted regions indicate the alignments in MFA and the correct alignments obtained using HS.}
  \label{fig:alignment}
\end{figure}

Figure \ref{fig:alignment} shows a sample Hindi waveform, its spectrogram and phone-level alignments obtained from different techniques. It is clearly seen that the alignments of TS are not correct. Although MFA has better alignments, the boundaries are more refined with HS. While HTS, DNN and E2E speech synthesis systems primarily learn the average statistical properties of phones, signal processing techniques rely on the acoustic properties of speech signals and do not require training. The acoustic features of syllables, such as the rising, steady state, and falling transitions, are well-known properties of syllables in speech. Restricting alignments to syllables yields accurate consonant boundaries too.
\subsection{Text processing}
\label{sec:testProc}
We convert the text to its phone-based representation using the unified parser for Indian languages \cite{arunTSD16}. The unified parser takes a word as input and applies relevant language-specific rules to generate the phone-based output in the common label set (CLS) representation \cite{ramani}. This output is further processed such that each phone is represented by a single character, as described in \cite{Prakash2019_MLCM}. Based on the duration information, each phone in the text is assigned a value equal to the number of frames. A comma is included in the text wherever the aligner has predicted a short pause ({\it sp}). Additional symbols ``\$'' and period ``.'' are included for beginning and end silence regions ({\it SIL}) in the audio (if present), respectively. 



\subsection{E2E training}

The E2E system considered in this work has a 2-stage pipeline: (1) text to mel-spectrogram conversion using a Transformer-based encoder-decoder FastSpeech2 architecture \cite{fastspeech2}, and (2) speech reconstruction using the HiFi-GAN vocoder \cite{hifigan_NEURIPS2020}. 

In FastSpeech2, the text is converted to phone embeddings which are then passed through a series of feed-forward Transformer (FFT) blocks to generate the phone hidden sequence. The phone hidden sequence is then expanded to match the length of the mel-spectrogram sequence based on the duration information. Then the expanded phone sequence is passed through another set of FFT blocks at the decoder to generate mel-spectrogram frames. Pitch and energy embeddings are added to the phone hidden state to provide more variance information. During training, the phone durations are obtained from a teacher model (such as Tacotron2) or an external aligner. Pitch and energy values are extracted from the ground-truth audio files. Duration, pitch and energy predictors are trained and optimized with mean square error (MSE) loss. During synthesis, these prosodic features are predicted by the network.





HiFi-GAN is a GAN-based vocoder capable of producing high-fidelity speech from mel-spectrograms \cite{hifigan_NEURIPS2020}. It is a non-autoregressive vocoder that models periodic patterns in speech audio. HiFi-GAN has a smaller footprint size and a higher synthesis speed compared to most neural vocoders.


\section{Experiments and results}
\label{sec:experiments}
\vspace{-0.3cm}

Systems are trained for 13 Indian languages from the open-source Indic TTS database \cite{ArunResources2016}\footnote{\url{https://www.iitm.ac.in/donlab/tts/database.php}}. The languages are Assamese, Bengali, Bodo, Gujarati, Hindi, Kannada, Malayalam, Manipuri, Marathi, Odia, Rajasthani, Tamil and Telugu. These languages span eight written scripts and three different language families-- Indo-Aryan, Dravidian and Sino-Tibetian. Separate systems are trained for male and female voices, except for Bodo, which has only a female dataset. Each dataset consists of about 10 hours of data spoken by a single person. A total of 25 TTS systems are trained using the proposed approach.

Audio files are downsampled to 22.05 kHz to ensure uniformity in the feature extraction part. The text is processed as described in Section \ref{sec:testProc}. $10\%$ of the TTS data is considered the validation set. For training Tacotron2 and FastSpeech2 models, the ESPNet (v2) toolkit was used \cite{espnet}, with the default parameters. HiFi-GAN v1 models were trained using an open-source code\footnote{\url{https://github.com/jik876/hifi-gan}}. The hybrid segmentation code\footnote{\url{www.iitm.ac.in/donlab/tts/hybridSeg.php}}, implemented using HTK \cite{htk} and Kaldi \cite{Povey_ASRU2011_2011} toolkits, was used. Training time for FastSpeech2 was 1.5-2 days for each dataset on 2 NVIDIA A100 40GB GPUs.



We first perform all experiments with the Hindi male dataset as proof of concept. The evaluation includes subjective and objective measures, alignment accuracy and spectrogram analysis. Based on these results, we then present comparative results of the proposed approach for all languages with respect to the best systems in \cite{ai4bharat_TTS_2023}.

\vspace{-0.2cm}
\subsection{Comparison with different alignment techniques (full data)}
For the Hindi male dataset, three systems are trained (with full data) based on the alignments used-- (1) from Tacotron2 as the teacher model (TS), (2) with MFA, and (3) using hybrid HMM-GD-DNN alignments (HS). We first calculate mel-cepstral distortion (MCD) scores \cite{MCD}, which is an objective measure. MCD gives a measure of the cepstral distortion of a synthesised utterance with respect to a reference. For this, 50 additional ground-truth audio files recorded by the same Hindi male speaker are used. MCD scores corresponding to different systems are given in Table \ref{tab:full_obj}. It is seen that the performance of all three systems is comparable, and the differences in scores are statistically not significant ($p > 0.05$).

\begin{table}[h!]
\centering
\caption{MCD scores corresponding to Hindi male systems with full data}
\label{tab:full_obj}
\begin{tabular}{|l|c|c|c|}
\hline
\multicolumn{1}{|c|}{\textbf{Systems}} & \textbf{TS}               & \textbf{MFA}                & \textbf{HS}             \\ \hline
\textbf{MCD score}                                  & \multicolumn{1}{r|}{6.56} & \multicolumn{1}{r|}{6.61} & \multicolumn{1}{r|}{6.58} \\ \hline
\end{tabular}
\end{table}

Since humans are the end-users of this technology, we also conduct a modified pairwise comparison (PC) test to assess the comparative system performance. In the PC test, listeners are presented with a pair of audio files in random order of systems, and asked to give their preference. In addition, evaluators also rate the quality of the synthesised utterances on a scale of 1-5, 5 being the best. This is similar to the mean opinion score (MOS), except that the evaluators listen to each audio pair and then rate each utterance corresponding to a system. We refer to this score as \textit{comparative MOS}.

The test sentences for the subjective evaluation were selected from the web, ensuring coverage of different domains-- news, sports, entertainment, and technical lectures. In our experience, conducting long subjective evaluations leads to listener fatigue. Hence, each listener evaluated 10 audio pairs among the 20 audio pairs in the test. 14 native Hindi listeners participated in each PC test.

\begin{table}[h!]
\centering
\caption{PC test results: Hindi male systems with full data-- preference in $\%$ (comparative MOS)}
\label{tab:pc_full}
\begin{tabular}{|l|r|r|r|}
\hline
\multicolumn{1}{|c|}{\textbf{System pairs}} & \multicolumn{1}{c|}{\textbf{TS/MFA}} & \multicolumn{1}{c|}{\textbf{HS}} & \multicolumn{1}{c|}{\textbf{Equal}} \\ \hline
TS vs. HS                                   & 10.99 (3.81)                               & 51.65 (4.13)                            & 37.36                               \\ \hline
MFA vs. HS                                  & 7.69 (3.24)                                 & 61.54 (4.10)                           & 30.77                               \\ \hline
\end{tabular}
\end{table}

Results of the PC test comparing TS vs. HS and MFA vs. HS systems with full data are presented in Table \ref{tab:pc_full}. On average, the system with HS is preferred in more than 56\% of the cases, with an equal preference of 34\% across the competing systems. The difference in performance between the baseline and proposed systems is extremely statistically significant ($p < 0.05$). Surprisingly, the synthesised output using the TS model is still good, despite the poor alignments shown in Figure \ref{fig:alignment}. On further investigation, we find that the mistakes in alignments follow consistent patterns across various audio files and hypothesise that the duration prediction is accordingly learnt given enough training data.



\vspace{-0.2cm}
\subsection{Comparison with different alignment techniques (low-resource scenario)}
In the low-resource scenario, only 1 hour of TTS data is considered for obtaining alignments and training. Here, four systems are trained with 1 hour of Hindi male data-- (1) FastSpeech2 with alignments from Tacotron2 as the teacher model (TS), (2) FastSpeech2 with MFA, (3) FastSpeech2 with hybrid HMM-GD-DNN alignments (HS), and (4) VITS model with alignments from MAS. MCD scores corresponding to these systems are presented in Table \ref{tab:1hr_obj}. The FastSpeech2 system with TS does not train well as the training of the Tacotron2 (1 hour) teacher model failed due to lack of adequate data. This is reflected in its high MCD score. The cepstral distortion across MFA and HS systems is similar. The MCD score of the VITS model is the least. This is contrary to our expectation as the VITS model makes very obvious perceptual mistakes (such as confusing the ``h'' and ``a'' sounds in many instances). We hypothesise that the lower MCD score of VITS may be due to its ``cleaner'' synthesised audio as a result of complete text-to-wav training, compared to slightly more ``noisy'' audio synthesised by the 2-stage E2E pipeline.

\begin{table}[h!]
\centering
\caption{MCD scores corresponding to Hindi male systems with 1 hour data}
\label{tab:1hr_obj}
\begin{tabular}{|l|c|c|c|c|}
\hline
\multicolumn{1}{|c|}{\textbf{Systems}} & \textbf{TS}               & \textbf{MFA}                & \textbf{HS} & \textbf{VITS}            \\ \hline
\textbf{MCD score}                                  & \multicolumn{1}{r|}{10.97} & \multicolumn{1}{r|}{7.30} & \multicolumn{1}{r|}{7.21} & \multicolumn{1}{r|}{6.95} \\ \hline
\end{tabular}
\end{table}

The modified PC test is also conducted in the low-resource scenario. The TS system is excluded from this test due to its poorly synthesised audio. Each listener evaluated a set of 10 audio pairs out of 20 audio pairs in the test. 14 and 13 native listeners participated in the MFA vs. HS and VITS vs. HS tests, respectively. Results of the PC tests are presented in Table \ref{tab:pc_1hr}. In the MFA vs. HS test, although the preference for the HS system in the low resource scenario has reduced (in comparison to that in full data), the system still outperforms the MFA (1 hour) model. The HS system also outperforms the VITS model in the VITS vs. HS test. The difference in performance of systems is statistically significant ($p < 0.05$). The performance of the VITS model with a lower MCD score and lower subjective preference is consistent with the observations in \cite{ai4bharat_TTS_2023}.

\vspace{-0.5cm}

\begin{table}[h!]
\centering
\caption{PC test results: Hindi male systems with 1 hour data-- preference in $\%$ (comparative MOS)}
\label{tab:pc_1hr}
\begin{tabular}{|l|c|c|c|}
\hline
\multicolumn{1}{|c|}{\textbf{System pairs}} & \textbf{MFA/VITS}               & \textbf{HS}                & \textbf{Equal}             \\ \hline
MFA vs. HS                                  & \multicolumn{1}{r|}{16.48 (3.47)} & \multicolumn{1}{r|}{36.26 (3.84)} & \multicolumn{1}{r|}{47.25} \\ \hline

VITS vs. HS                                  & \multicolumn{1}{r|}{13.08 (2.88)} & \multicolumn{1}{r|}{60.77 (3.68)} & \multicolumn{1}{r|}{26.15} \\ \hline

\end{tabular}
\end{table}


\vspace{-0.7cm}

\subsection{Alignment accuracy}
\vspace{-0.7cm}

\begin{table}[h!]
\centering
\caption{Alignment accuracy}
\label{tab:diff}
\begin{tabular}{|l|l|l|}
\hline
\textbf{Alignment technique}                & \multicolumn{1}{c|}{MFA} & \multicolumn{1}{c|}{HS} \\ \hline
\textbf{Duration difference (in ms)} & 11.88                    & 4.40                    \\ \hline
\end{tabular}
\end{table}
Across experiments on both full and 1-hour data, we see that the MFA system is the closest common competing system to the proposed system (from both objective and subjective measures). Hence, we perform further comparative analysis across these systems. To check the accuracy of alignments obtained using MFA and HS alignment techniques on full data, we manually align 10 randomly chosen ground truth utterances of the Hindi male training data at the phone level. The average of absolute boundary differences with different alignments (of full data) is given in Table \ref{tab:diff}. It is clearly seen that HS provides more accurate alignments compared to MFA.


\subsection{Spectrogram analysis}

\begin{figure}[h!]
 \centering
 \includegraphics[width = \linewidth, height=6.5cm]{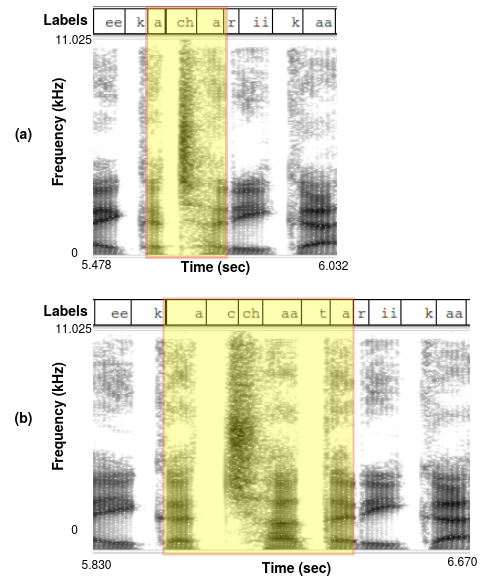}
 \vspace{-0.7cm}
   \caption{Spectrograms of synthesised utterances of Hindi male systems (with full data) using MFA (top) and HS (bottom) corresponding to the text ``eek acchaa tariikaa''.}
  \label{fig:spec_full}
\end{figure}

\begin{figure}[h!]
 \centering
 \includegraphics[width = \linewidth, height = 6.5cm]{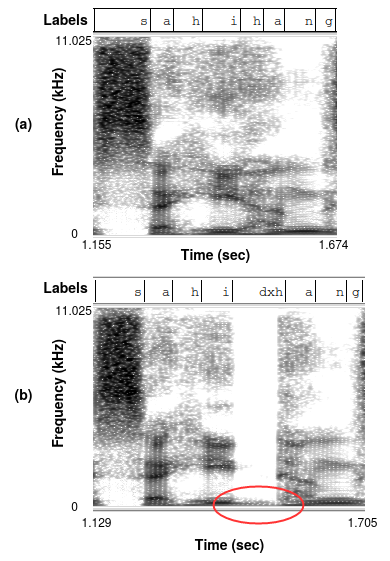}
   \vspace{-0.7cm}
   \caption{Spectrograms of synthesised utterances of Hindi male systems (with 1 hour data) using MFA (top) and HS (bottom) corresponding to the text ``sahi dxhang''.}
  \label{fig:spec_1hr}
\end{figure}

\begin{table*}[htb]
\footnotesize
\centering
\caption{Comparison of proposed systems and existing best systems for various Indian languages: Preference in \% (comparative MOS). The number of evaluators for each language is indicated next to the language.}
\label{tab:comp_ai4bharat}
\begin{tabular}{l|ccc|ccc}
\hline
\multicolumn{1}{c|}{\multirow{2}{*}{\textbf{Language}}} & \multicolumn{3}{c|}{\textbf{Male voice}}              & \multicolumn{3}{c}{\textbf{Female voice}}                \\ \cline{2-7} 
\multicolumn{1}{c|}{}                                   & \textbf{Proposed} & \textbf{Existing \cite{ai4bharat_TTS_2023}} & \textbf{Equal} & \textbf{Proposed} & \textbf{Existing \cite{ai4bharat_TTS_2023}} & \textbf{Equal} \\ \hline
Assamese (9)$^*$                                               & 34.72 \textbf{(3.15)}                 & 20.83 (2.95)                 & \textbf{44.45}              & \textbf{69.44 (3.72)}                 & 8.33 (2.69)                 & 22.23              \\ \hline
Bengali (16)                                               & \textbf{71.09 (3.92)}                 & 10.94 (3.08)                 & 17.97              &  \textbf{72.66 (4.14)}                & 12.50 (3.23)                 & 14.84              \\ \hline
Bodo (16)                                                   & --                 & --                 & --              & \textbf{58.93 (3.93)}                & 19.64 (3.19)                & 21.43             \\ \hline
Gujarati (13)                                               & \textbf{45.19 (3.75)}                 & 37.50 (3.54)                 & 17.31              & \textbf{80.77 (4.17)}                 & 7.69 (3.20)                 & 11.54              \\ \hline
Hindi (28)                                                  & \textbf{57.14 (3.91)}                 & 28.12 (3.56)                 & 14.74              & \textbf{69.64 (4.26)}                 & 9.38 (3.57)                 & 20.98              \\ \hline
Kannada (11)                                               & \textbf{70.45 (4.17)}                 & 12.50 (3.22)                 & 17.05              & \textbf{48.86 (4.20)}                 &  11.36 (3.69)                & 39.78              \\ \hline
Malayalam (20)                                              & \textbf{64.38 (4.04)}                 & 14.38 (3.26)                 & 21.24              & \textbf{43.75 (3.79)}                 & 32.50 (3.65)                 & 23.75              \\ \hline
Manipuri (6)$^*$                                              & \textbf{52.08 (2.83)}
& 31.25 (2.51)                 & 16.67              & 37.50 (2.68)                & \textbf{41.67 (2.72)}                 & 20.83              \\ \hline
Marathi (21)                                               & \textbf{77.98 (4.21)}                 & 11.31 (3.16)                 & 10.71              & \textbf{76.78 (4.02)}                 & 14.88 (3.16)                 & 8.34              \\ \hline
Odia  (8)$^*$                                                 & \textbf{68.75 (3.55)}                 & 25.00 (2.86)                 & 6.25              & \textbf{59.38 (3.19)}                 & 29.69 (2.90)                 & 10.93              \\ \hline
Rajasthani (2)$^*$                                             & \textbf{56.25 (3.84)}                 & 12.50 (3.53)                 & 31.25              & \textbf{93.75 (4.47)}                & 0 (3.78)                 & 6.25              \\ \hline
Tamil (22)                                                 & \textbf{68.18 (4.16)}                 & 21.02 (3.54)                 & 10.80              &  \textbf{55.11 (3.95)}                & 28.41 (3.61)                 & 16.48              \\ \hline
Telugu (15)                                                & \textbf{51.67 (3.87)}                 & 29.17 (3.47)                 & 19.16              & \textbf{84.17 (3.73)}                 & 8.33 (2.69)                 & 7.50              \\ \hline
\textbf{Average}                                                & \textbf{59.82 (3.78)}                 & 21.21 (3.22)                 & 18.97              & \textbf{65.44 (3.86)}                 & 17.26 (3.24)                 & 17.30              \\ \hline
\multicolumn{6}{l}{\small $^*$Results are indicative and not conclusive due to the lack of evaluators.} \\
\end{tabular}
\end{table*}

Figures \ref{fig:spec_full} and \ref{fig:spec_1hr} show spectrograms of utterances synthesised by the MFA and HS systems. Consider the highlighted regions in Figure \ref{fig:spec_full}. The HS system correctly generates audio corresponding to the text ``eek acchaa tariikaa''. However, the MFA system misses a few sounds, and the utterance is perceived as ``eek chariikaa''. The short vowel ``a'' at the beginning of ``acchaa'' is hardly perceived (probably due to having a short duration), and the voiceless stop consonant ``t'' in ``tariikaa'' appears to be replaced by the aspirated affricate ``ch''. In Figure \ref{fig:spec_1hr}, the aspirated voiced stop consonant ``dxh'' in ``dxhang'' is missed by the MFA system, while it is uttered correctly by the HS system (as evidenced by the voice bar).

Our observations from the experiments conducted so far are summarised here. Although the alignments from the teacher model are poor (but mostly consistent) in many places, the FastSpeech2 student model still learns to generate good-quality speech, given enough amount of training data. But more accurate alignments are required to further improve the pronunciation of sounds in the generated output, especially in low-resource scenarios. In this context, signal processing cues, such as GD and SBSF, in tandem with deep learning techniques, aid in providing accurate alignments. It is to be noted that the accuracy of alignments also depends on the accuracy of transcriptions in correspondence with the training utterances and the accuracy of the word-to-phone lexicon.



 \vspace{-0.35cm}
\subsection{Comparison with existing best models for Indian languages}
 \vspace{-0.1cm}
Encouraged by the results of the experiments conducted, FastSpeech2 based systems with the hybrid HMM-GD-DNN alignments are trained for 13 Indian languages, with male and female voices. These systems are compared with the existing best TTS models available for Indian languages \cite{ai4bharat_TTS_2023}\footnote{\url{https://models.ai4bharat.org/#/tts}}. It is to be noted that both sets of systems are trained on the IndicTTS database \cite{ArunResources2016}.  The test sentences are selected from the web, covering various domains-- news, sports, entertainment, and technical lectures. However, for Bodo, Manipuri and Rajasthani, sentences from the eval set (not seen during training) have been considered, as it was difficult to find out-domain sentences in those languages. 



The modified PC test is conducted to evaluate the comparative performance of these systems. Totally, 187 listeners participated in the evaluations. The number of native speakers for each language is given in Table \ref{tab:comp_ai4bharat}. Rajasthani systems were evaluated by only 2 listeners, as it was very difficult to find native speakers of that language. Results of languages with less than 10 evaluators have been mentioned as indicative rather than conclusive (denoted by a $^*$). Each evaluator evaluated 8 audio pairs each for male and female TTS system comparisons. 

It is clearly seen from Table \ref{tab:comp_ai4bharat} that the proposed systems perform better than the systems in \cite{ai4bharat_TTS_2023} in all cases, except Manipuri female, where the degradation is marginal. The difference in performance of 18 models out of 25 is extremely statistically significant ($p < 0.05$). For the following systems, the difference in scores is not statistically significant: Assamese (male), Gujarati (male), Malayalam (female), Manipuri (male, female), Odia (female), Rajasthani (male).

As seen in Table \ref{tab:comp_ai4bharat}, the performance of systems across languages varies. The TTS synthesis quality is limited by the quality of the TTS data used for training. Factors such as voice timbre, syllable rate, speaking speed, enunciation, phone coverage and completeness of the text, accuracy of transcriptions impact the output synthesis quality. As a result, systems corresponding to some of the languages perform better than others.







 \vspace{-0.65cm}
\section{Conclusion}
\label{sec:conclusion}
\vspace{-0.4cm}

In this work, we have built good quality TTS systems for 13 Indian languages by seamlessly integrating signal processing cues in E2E system building. We have seen how accurate phone boundaries for the training data have led to better duration modelling, and consequently to better synthesis. We can further reduce the amount of data (to 30 minutes, 15 minutes and so on) to stress test the systems trained with different alignments. This work can be further extended to other prosodic parameters, namely, stress and pitch. Similar ideas can be explored in the context of other direct text-to-speech E2E systems, as signal processing primarily depends on the acoustic characteristics of the speech signal, is agnostic to text, and is complementary to the model used.




\vspace{-0.3cm}
\section{Acknowledgement}
\vspace{-0.4cm}

We thank the Ministry of Electronics and Information Technology (MeitY), Government of India (GoI), for funding the project ``Speech Technologies in Indian Languages'' (SP/21-22/1960/CSMEIT/003119). Special thanks to John Wesly for helping out in the subjective evaluations.

\bibliographystyle{IEEEbib}
\bibliography{refs}

\end{document}